\documentclass[conference]{IEEEtran}
\pdfoutput=1
\usepackage[noadjust]{cite}
\ifCLASSINFOpdf
	\usepackage[pdftex]{graphicx}
\else

\fi

\usepackage{amsmath}
\usepackage{array}

\usepackage{url}

\usepackage{listings}
\usepackage{multirow}

\usepackage[usenames, dvipsnames]{color}
\usepackage{fancyvrb}

\definecolor{mypink1}{rgb}{0.858, 0.188, 0.478}

\begin{document}

\title{Anonymization of System Logs for \\Privacy and Storage Benefits}

\author{\IEEEauthorblockN{Siavash Ghiasvand}
\IEEEauthorblockA{Center for Information Services\\ and High Performance Computing\\
Technical University of Dresden, Germany\\
Email: siavash.ghiasvand@tu-dresden.de}
\and
\IEEEauthorblockN{Florina M. Ciorba}
\IEEEauthorblockA{Department of Mathematics\\ and Computer Science\\
University of Basel, Switzerland\\
Email: florina.ciorba@unibas.ch}}

\maketitle

\begin{abstract}
System logs constitute valuable information for analysis and diagnosis of system behavior. 
The size of parallel computing systems and the number of their components steadily increase. 
The volume of generated logs by the system is in proportion to this increase. 
Hence, long-term collection and storage of system logs is challenging. 
The analysis of system logs requires advanced text processing techniques. 
For very large volumes of logs, the analysis is highly time-consuming and requires a high level of expertise. 
For many parallel computing centers, outsourcing the analysis of system logs to third parties is the only affordable option. 
The existence of sensitive data within system log entries obstructs, however, the transmission of system logs to third parties. 
Moreover, the analytical tools for processing system logs and the solutions provided by such tools are highly system specific. 
Achieving a more general solution is only possible through the access and analysis system of logs of multiple computing systems. 
The privacy concerns impede, however, the sharing of system logs across institutions as well as in the public domain.  
This work proposes a new method for the anonymization of the information within system logs that employs de-identification and encoding to provide sharable system logs, with the highest possible data quality and of reduced size. 
The results presented in this work indicate that apart from eliminating the sensitive data within system logs and converting them into shareable data, the proposed anonymization method provides 25\% performance improvement in post-processing of the anonymized system logs, and more than 50\% reduction in their required storage space.
\end{abstract}

\IEEEpeerreviewmaketitle

\section{Introduction}
\label{sec:introduction}
System logs are valuable sources of information for the analysis and diagnosis of system behavior.
The size of computing systems and the number of their components, continually increase.
The volume of generated system logs (hereafter, syslogs) is in proportion to this increase.
The storage of the syslogs produced by large parallel computing systems in view of their analysis requires high storage capacity.
Moreover, the existence of sensitive data within the syslogs raises serious concerns about their storage, analysis, dissemination, and publication.
The anonymization of syslogs is a means to address the second challenge.
During the process of anonymization, the sensitive information will be eliminated while the remaining data is considered as \emph{cleansed data}.
Applying anonymization methods to syslogs to cleanse the sensitive data before storage, analysis, sharing, or publication, reduces the usability of the anonymized syslogs for further analysis.
After a certain degree of anonymization, the cleansed syslog entries lose semantic and only remain useful for statistical analysis, such as time series and distributions.
At this stage, it is possible to transform long syslog entries into shorter strings.
Reducing the length of cleansed and semantic-less syslog entries significantly reduces the required storage capacity of syslogs and addresses the storage challenge mentioned earlier.
Shortening the log entries' length reduces their processing complexity and, therefore, improves the performance of further analysis on syslogs.

\begin{figure}[!t]
	\centering
	\includegraphics[width=2.5in]{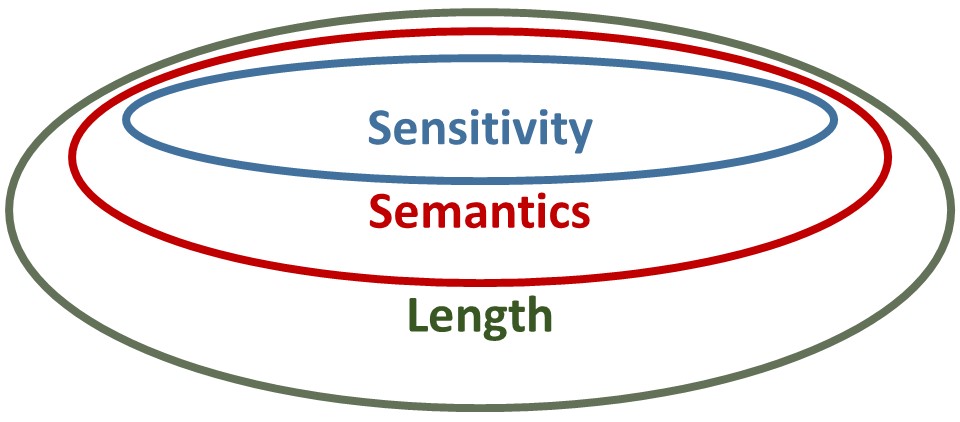}
	\caption{The sensitivity, semantic, and length of terms in syslog entries and their relation. Each term of a syslog entry has a non-zero length. Terms may or may not have semantic. A term with semantic may or may not also be sensitive.}
	\label{fig:venn}
\end{figure}

In this work, we address the trade-off between the sensitivity and the usefulness of the information in anonymized syslogs.
It is important to note that the sensitivity and the semantic of syslogs are relative terms.
Each data item (or term) in a syslog entry, depending on policies of the computing system it originates from, may or may not be considered sensitive data.
The same degree of relativity applies to the semantic of a syslog entry data item.
Depending on the chosen data analysis method, the semantic of syslogs can be assessed as rich or poor.
Even though the classification of each term as sensitive or as semantic is related to the policies of computing centers, the final assessment of sensitivity and semantic has a binary value of \emph{true} (1) or \emph{false} (0).
Therefore, every single term in a syslog entry can only be sensitive or nonsensitive, e.g., a username.
Fig.~\ref{fig:venn} illustrates the relation between the sensitivity, the semantic and the length of the syslog terms.

\begin{figure}[!t]
	\centering
	\includegraphics[width=0.5\textwidth]{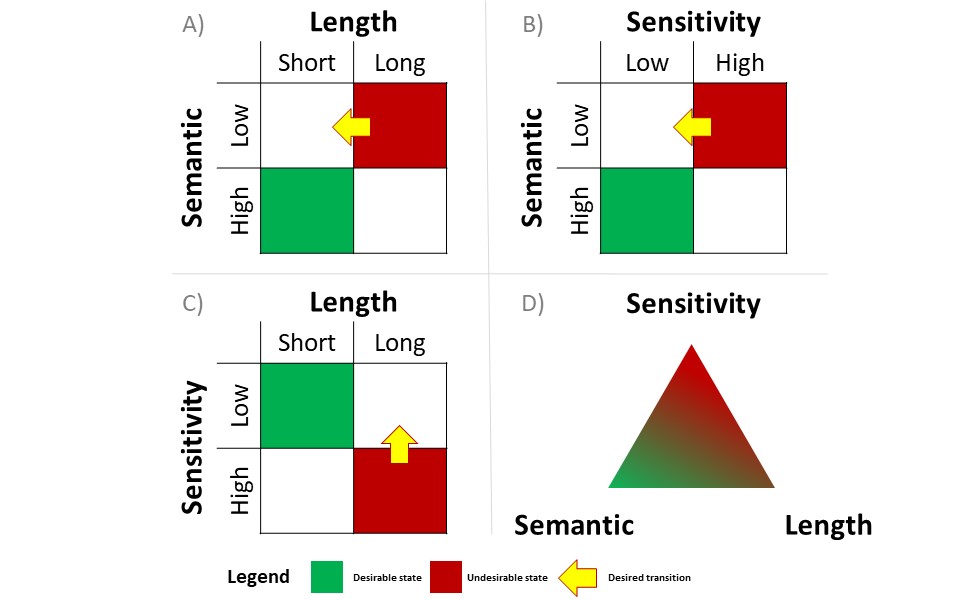}
	\caption{Trade-off scenarios between the semantic, sensitivity, and length of a system log entry.
		Each of the A), B), and C) illustrations depicts the four possible states of a syslog entry based on its sensitivity, semantic, and length.
		These states are not limited to syslog entires alone and can be generalized to higher granularities of syslog data.
		The green state denotes the best situation, while the red state is most undesirable.
		In red states, the intention is transition to the state shown by the yellow arrow in the trade-off diagrams.
		The trade-off triangle in illustration D) shows the trade-off between the three parameters (sensitivity, semantic, length) in a single unified view.}
	\label{fig:tradeoff}
\end{figure}

A triple trade-off exists between sensitivity, semantic, and length of a syslog entry.
Fig.~\ref{fig:tradeoff} schematically illustrates this trade-off, regardless of the system policies and syslog analysis methods in use.
The illustration shows that a syslog entry can be in four distinct states.
Green color states denote best conditions while red color states denote undesirable conditions.
White color states represent neutral conditions.
Under undesirable conditions, the approach taken in this work is to transition from the red state to one of the white states.
The yellow arrows indicate this in Fig.~\ref{fig:tradeoff}.
Increasing the semantic of a syslog entry is not possible.
Therefore, the remaining possibilities are either decreasing the sensitivity or reducing the length of the syslog entry.

Data, in general, has a high quality when it is ”fit for [its] intended uses in operations, decision making, and planning”~\cite{quality}. 
The syslog entries represent the data in this work and several parameters affect their quality. 
To measure and maximize this quality, a utility function called \emph{quality} ($Q_E$) is defined as the relation between sensitivity, semantic, length, and usefulness of syslog entry \emph{$E$}. 
The goal of this work is to maintain the \emph{quality} ($Q_E$) of all syslog entries, by pushing the parameters mentioned above toward their best possible values, when the computing system policies degrade this quality.
The main contribution of this work is in introducing a new approach for anonymization that employs de-identification and encoding to provide shareable system logs, with the highest possible data quality and of reduced size.

The remainder of this work is organized as follows.
In Section~\ref{sec:relatedwork} the background and current state of the art are discussed. 
The proposed approach is described in Section~\ref{sec:approach}, and the methodology and technical details are provided in Section~\ref{sec:methodology}.
After explaining the results of the current work in Section~\ref{sec:results}, the conclusion and future work directions are discussed in Section~\ref{sec:conclusion}.

\section{Related work}
\label{sec:relatedwork}
In July 2000, the European Commission adopted a decision recognizing the "Safe Harbor Privacy Principles"~\cite{safe}.
Based on the "Safe Harbor" agreement, eighteen personal identifiers should be eliminated from the data before its transmission and sharing.
"Safe Harbor" was originally designed to address the privacy of healthcare-related information.
However, its principles are also taken into account for other types of information.

Later, in March 2014, European Parliament approved the new privacy legislation.
According to this regulations, \emph{personal data} is defined as "any information relating to an identified or identifiable natural person (`data subject.')"~\cite{GDPR}. 
This information must remain private to ensure a person's privacy.
Based on this definition, syslog entries contain numerous terms which represent personal data and must, therefore, be protected.

Protection of personal data in syslog entries can be attained via various approaches; the most common ones are encryption and de-identification.
Encryption reduces the risk of unauthorized access to personal data. 
However, the encrypted syslog entries cannot be freely used or shared in the public domain.
The risk of disclosure of the encryption-key also remains an important concern.
In contrast, de-identification eliminates the sensitive data and only preserves the nonsensitive (cleansed) data. 
As such, de-identification provides the possibility of sharing de-identified information in the public domain.
The de-identified data may turn out to no longer be of real use.

Pseudonymization and anonymization are two different forms of de-identification.
In pseudonymization, the sensitive terms are replaced by dummy values to minimize the risk of disclosure of the \emph{data subject} identity.
Nevertheless, with pseudonymization the \emph{data subject} can potentially be re-identified by some additional information~\cite{sweeney}.
Anonymization, in contrast, refers to protecting the user privacy via irreversible de-identification of personal data.

Several tools have been developed to address the privacy concerns of using syslog information.
Most of these tools provide log encryption as the main feature, while certain such tools also provide de-identification as an additional feature.
Syslog-ng and Rsyslog are two open-source centralized logging infrastructures that provide \emph{out of the box} encryption and message secrecy for syslogs, as well as de-identification of syslog entries~\cite{syslog-ng1, rsyslog2}.
Both tools provide a \emph{pattern database} feature, which can identify and rewrite personal data based on pre-defined text patterns.
Logstash~\cite{logstash} is another open-source and reliable tool to parse, unify, and interpret syslog entries.
Logstash provides a text filtering engine which can search for the text patterns in live streams of syslog entries and replace them with predefined strings~\cite{Sanjappa2017}.
In addition to the off-line tools, such as Syslog-ng and Logstash, there is a growing number of on-line tools, e.g., Loggy~\cite{Loggy}, Logsign~\cite{Logsign}, and Scalyr~\cite{Scalyr}, that offer a comprehensive package of syslog analysis services.
The existence of sensitive data in the syslogs, barricades the usage of such services.

Alongside these industrial-oriented tools, several research groups have developed scientific-oriented toolkits to address the syslog anonymization challenge.
eCPC toolkit~\cite{eCPC}, sdcMicro~\cite{sdcMicro}, TIAMAT~\cite{TIAMAT}, ANON~\cite{ANON}, UTD Anonymization Toolbox~\cite{UTD}, and Cornell Anonymization Toolkit~\cite{Xiao} are selected examples of such toolkits.
These tools apply various forms of \emph{k-anonymity}~\cite{sweeney} and \emph{l-diversity}~\cite{diversity} to ensure data anonymization.
Achieving an optimal \emph{k-anonymity} is an NP-hard problem~\cite{Meyerson}.
Heuristic methods, such as k-Optimize, can provide effective results~\cite{Bayardo}.

The main challenges of using existing anonymization approaches, in general, are:
(1)~The quality of the anonymized data dramatically degrades, and
(2)~The size of the anonymized syslogs remains almost unchanged.
The industrial-oriented approaches are unable to attain full anonymization at micro-data~\cite{sweeney} level.
Even though scientific-oriented approaches can guarantee a high level of anonymization, they are mainly not capable of applying effective anonymization in an online manner.
Certain scientific-oriented methods, such as~\cite{Rath}, which can effectively anonymize online streams of syslogs, need to manipulate log entries at their origin~\cite{Rath-tool}.

The anonymization approach proposed in this work is distinguished from existing work through the following features: 
(1)~Ability to work with streams of syslogs without modification of the syslog origin; 
(2)~Preservation of the highest possible quality of log entries; and 
(3)~Reduction of the syslogs storage requirements, whenever possible.
\section{Proposed Approach}
\label{sec:approach}
Computing systems can generate system logs in various formats.
RFC5424 proposes a standard for the syslog protocol which is widely accepted and used on computing systems~\cite{syslog}.
According to this protocol, all syslog entries consist of two main parts: a \emph{timestamp} and a \emph{message}.
In addition to these two main parts, there are optional parts, such as \emph{system tags}.
Let us consider the following sample syslog entry $E_1$: "\fontsize{9pt}{10pt}\verb|1462053899 Accepted publickey for Siavash from|
\verb|4.3.2.1|".
In this entry, "\fontsize{9pt}{10pt}\verb|1462053899|\fontsize{10pt}{10pt}" is the \emph{timestamp} and the rest of the line "\fontsize{9pt}{10pt}\verb|Accepted publickey for|
\verb|Siavash from 4.3.2.1|\fontsize{10pt}{10pt}" is the \emph{message}.
In the \emph{message} part, the terms \verb|Accepted|, \verb|publickey|, \verb|for|, and \verb|from| are \emph{constant} terms, while \verb|Siavash|, and \verb|4.3.2.1| are \emph{variable} terms, in the sense that for the above constant terms, the user name and IP can vary among users and machines.

The goal of this work, described earlier in Section~\ref{sec:introduction}, is to preserve the \emph{quality} of syslog entries throughout the anonymization process.
To achieve this goal,
(1)~The \emph{variable} terms in the syslog entries are divided into 3 groups:  \emph{sensitive}, \emph{meaningful} (those that have a semantic), and \emph{semantic-less} terms.
(2)~The \emph{sensitive} terms are eliminated to comply with the privacy policies.
(3)~The \emph{semantic-less} terms are replaced with predefined constants to reduce the required storage.
(4)~Following the anonymization steps (2) and (3) above, every syslog entry that does not have any additional \emph{variable} terms, is mapped to a hash-key, via a collision-resistant hash function. This step is called \emph{encoding}.
(5)~The quality of the remaining syslog entries is measured with a utility function.
(6)~When it is revealed that removing a \emph{meaningful} term from the syslog entry improves the quality of syslog, that particular term is replaced with a predefined constant.
(7)~The remaining processed syslog entries that do not contain additional \emph{variable} terms, are mapped into hash-keys (similar to step (4) above).
(8)~Upon completion of steps (4) and (7), the hash-key codes can be optimized based on their frequency of appearance.
The preliminary results of analyzing the syslogs of a production HPC system called Taurus\footnote{https://doc.zih.tu-dresden.de/hpc-wiki/bin/view/Compendium/SystemTaurus} using the proposed approach shows up to 95\% reduction in storage capacity~\cite{ISC17}.
An interactive demonstration of the use of this anonymization approach on a sample syslog is provided online~\cite{demo}.

In the proposed approach, regular expressions are used for the automatic detection of variable terms within syslog entries.
Categorization of automatically detected terms into \emph{sensitive} and/or \emph{meaningful} is performed based on the information in Table~\ref{tab:sense-semantic}.
This information is inferred from the policies and conditions of the host high-performance computing system.
Automatically detected variable terms which do not belong to any of the \emph{sensitive} and \emph{meaningful} categories are considered as \emph{semantic-less}.
A variable length hash algorithm is used to encode the syslog entries.
The encoding step is described in greater details at the end of this section.

Table~\ref{tab:regex} contains fifteen main regular expressions (out of thirty-eight) which are used to detect variable terms in syslog entries.
The order of their application is significant since certain patterns are subsets of other patterns.
Even though most variables can be detected with these regular expressions, in an unlikely case of similarity between variables and constants, the regular expression may not be able to differentiate between constants and variables correctly.
For example the username \emph{panic} may be misinterpreted as a constant value like \emph{kernel panic}.
In such scenarios, the undetected variables are considered as constants (or vice versa) and  appear as a new event pattern.
Encoding event patterns in the final step of the proposed approach guarantees the highest attainable level of anonymization.

\begin{table}[h]
	\caption{
	The variable terms in the Taurus syslogs can be detected with thirty-eight regular expressions. Out of those, the main fifteen machine-independent regular expressions are shown here. These can be used to identify the variable terms within syslogs from any computing system.}
	\label{tab:regex}
	\begin{tabular}{ll}
		\textbf{Variable type}&Regular expression\\
		Path & \fontsize{7.2pt}{10pt}\Verb|([\(\s\,\>\:\=])([\/][a-z0-9_\.\-\:]*)+| \\
		Version & \fontsize{7.2pt}{10pt}\Verb|([\w\.\-]+x86_64)| \\
		Email & \fontsize{7.2pt}{10pt}\Verb|([a-z0-9_\-\.]+@([a-z0-9_-]+\.)+[a-z]+)| \\
		DateTime & \fontsize{7.2pt}{10pt}\Verb|(\d{4}-\d{2}-\d{2})T(\d{2}:\d{2}:\d{2})| \\
		IPv4 & \fontsize{7.2pt}{10pt}\Verb|(\d+\.\d+\.\d+\.\d+)| \\
		Port & \fontsize{7.2pt}{10pt}\Verb|([\W])(port \d+)| \\
		Parameter & \fontsize{7.2pt}{10pt}\Verb|(\$[a-z0-9_]+)| \\
		URID & \fontsize{7.2pt}{10pt}\Verb|(uid=[\w\-]+)| \\
		User & \fontsize{7.2pt}{10pt}\Verb|(for )((user\ )*[a-z0-9_-]+)| \\
		Library & \fontsize{7.2pt}{10pt}\Verb|([a-z0-9_\-]+\.so(\.\d*)*)| \\
		Hardware address & \fontsize{7.2pt}{10pt}\Verb|(0[x][a-f0-9]+\-0[x][a-f0-9]+)| \\
		Hex Number & \fontsize{7.2pt}{10pt}\Verb|(0[x][a-f0-9]+)| \\
		Percentage & \fontsize{7.2pt}{10pt}\Verb|(\d+\.*[\d]*\%)| \\
		Serial number & \fontsize{7.2pt}{10pt}\Verb|((\s)([a-f0-9\.\-]+\:)+(\s))| \\
		Size & \fontsize{7.2pt}{10pt}\Verb|([^a-z0-9])(\d+[bkmg])([^a-z0-9])| \\
	\end{tabular}
\end{table}

As the first step, the \emph{quality} of syslog entries needs to be quantified.
The product of four characteristics of syslog entries defines the syslog entry quality: (1)~sensitivity, (2)~semantic, (3)~length, and (4)~usefulness.
To render uniform the impact of all characteristics, their significance is normalized in the range of 0 to 1, and the negative parameters are replaced with their reverse positive counterparts.
Therefore, the effective parameters are \emph{nonsensitivity}, \emph{semantic}, \emph{reduction}, and \emph{usefulness}.
The \emph{nonsensitivity} parameter of a syslog entry can take any value in the range of 0 (most sensitive) to 1 (most nonsensitive) denoting the best value.
The parameter \emph{semantic} can also take any value in the range of 0 (least semantic) to 1 (highest semantic), with 1 representing a highly relevant syslog entry.

The length of syslog entries can be interpreted as the size of syslog entry.
The \emph{reduction} of syslog entries size may also take any value in the range of 0 (no reduction) to 1 (most reduction).
Size \emph{reduction} can be achieved via any general lossy or lossless compression algorithm.
When the applied compression method does not change the semantic and sensitivity of syslog entries, it is considered as lossless (from the perspective of this work).
If the chosen compression method modifies the semantic or sensitivity of syslog entries, in the context of this work, it is taken as an additional level of anonymization rather that compression.
A careful consideration of various effective compression algorithms, including Brotli, Deflate, Zopfli, LZMA, LZHAM, and Bzip2, revealed that in affordable time, compression could reduce the data size to 25\% of its original size.
Therefore the \emph{reduction} of syslogs ranges between 0.75 to 1, where 1 indicates 100\% compression and is practically impossible to reach.
Every time that a compressed syslog entry is processed, the decompression process imposes an additional performance penalty on the host system.
Therefore, the proposed approach in this work uses an encoding algorithm instead of compression algorithms which demand a decompression before accessing the compressed data.
The encoded data can be accessed and used without pre-processing.

Unlike the previous three parameters, the fourth parameter, \emph{usefulness}, is boolean and takes 0 or 1 as values.
The value of 0 or 1 for \emph{usefulness} denotes that a syslog entry in its current from cannot or can be used for a specific type of analysis, respectively.
\vspace{-0.5em}
\begin{equation}
\label{quality}
\begin{split}
Q_E & = U_E * (n*N_E) * (s*S_E) * (r*R_E)
\end{split}
\end{equation}

Equation~(\ref{quality}) quantifies the \emph{quality}~($Q_E$) of a syslog entry $E$ as a product of its nonsensitivity~($N_E$), semantic~($S_E$), reduction~($R_E$), and usability~($U_E$).
The coefficients, $n$, $s$, and $r$ indicate the importance of nonsensitivity, semantic, and reduction for a specific computing system.
The default value for $n$, $s$, and $r$ is 1.
The value of 0 for usability results in a 0-quality syslog entry and disqualifies the current syslog entry from further analysis.
As explained earlier, regardless of system conditions and policies, a \emph{reduction} rate of 75\% is always achievable~\cite{Comp1, Comp2, Comp3, Comp4}.
Therefore, the \emph{quality} of a raw syslog entry is calculated using Equation~(\ref{quality_raw}).
\vspace{-0.5em}
\begin{equation}
\label{quality_raw}
\begin{split}
Q_E & = 1 * (1 * N_E) * (1 * S_E) * (1 * 0.75)
\end{split}
\end{equation}
 
The sensitivity of each syslog entry term is defined based on the policies set up by the computing system administrators.
Assume that Table~\ref{tab:sense-semantic} indicates the sensitivity and the semantic of syslog entry terms of a computing system.
The severity degree of each term's sensitivity varies from 0 to 10.
This degree is only used to give priority to the individual anonymization steps.
Each syslog entry term can be sensitive (Y) or nonsensitive (N).
Therefore, in this section, only the boolean sensitivity indicator (Y/N) is considered to denote the sensitivity of each syslog entry term.
The same assumptions hold for the semantic of each syslog entry term.
Accordingly, each term can be \emph{with} or \emph{without} semantic.
The semantic of each term can be judged from 3 sources.
(1)~Every sensitive term is also semantic.
(2)~Every semantic term is marked with "Y" in the semantic table (Table~\ref{tab:sense-semantic}),
(3)~All terms not included in the semantic table nor marked with "Y" therein simply have length, are nonsensitive, and have no semantic.

Table~\ref{tab:syslog_sample} indicates the sensitivity and semantic of each term from the message part of the sample syslog entry based on information from Table~\ref{tab:sense-semantic}.

\begin{table}[h]
	\caption{Classification of syslog entry terms into sensitive and/or semantic.
	Severity denotes the importance of the characteristics for the respective terms.}
	\label{tab:sense-semantic}
	\begin{tabular}{@{}l|@{ }c@{ }|@{ }c@{}}
		\textbf{Term}&\textbf{Sensitivity}&\textbf{Severity}\\
		\hline
		\fontsize{7.2pt}{10pt}\Verb|User Name|  &Y &10 \\
		\fontsize{7.2pt}{10pt}\Verb|IP Address| &Y &08 \\
		\fontsize{7.2pt}{10pt}\Verb|Port Number|&Y &01 \\
		\fontsize{7.2pt}{10pt}\Verb|Node Name|  &Y &03 \\
		\fontsize{7.2pt}{10pt}\Verb|Node ID|    &Y &03 \\
		\fontsize{7.2pt}{10pt}\Verb|Public Key| &Y &10 \\
		\fontsize{7.2pt}{10pt}\Verb|App Name|   &N &00 \\
		\fontsize{7.2pt}{10pt}\Verb|Path / URL| &N &00 \\
	\end{tabular}
	\hspace{0.25cm}
	\begin{tabular}{@{}l|@{ }c@{ }|@{ }c@{}}
	\textbf{Term}&\textbf{Semantic}&\textbf{Severity}\\
	\hline
	\fontsize{7.2pt}{10pt}\Verb|accept*|		&Y &07\\
	\fontsize{7.2pt}{10pt}\Verb|reject*| 		&Y &10\\
	\fontsize{7.2pt}{10pt}\Verb|close*|		&Y &08\\
	\fontsize{7.2pt}{10pt}\Verb|*connect*|	&Y &09\\
	\fontsize{7.2pt}{10pt}\Verb|start*|		&Y &02\\
	\fontsize{7.2pt}{10pt}\Verb|*key*|		&Y &01\\
	\fontsize{7.2pt}{10pt}\Verb|session|		&Y &07\\
	\fontsize{7.2pt}{10pt}\Verb|user*|		&Y &05\\	
	\end{tabular}
\end{table}

\begin{table}[h]
	\caption{A sample syslog entry. Sensitive and meaningful terms are marked with "Y" in the respective rows.}
	\label{tab:syslog_sample}
	\begin{tabular}{l|*{6}{c}}
		\textbf{Message}&Accepted&publickey&for&Siavash&from&4.3.2.1\\
		\textbf{Sensitive}&-&-&-&Y&-&Y\\
		\textbf{Semantic}&Y&Y&-&Y&-&Y\\
	\end{tabular}
\end{table}

The nonsensitivity ($N_E$) of a syslog entry $E$ is defined as $\frac{\text{Number\ of\ nonsensitive\ terms\ in\ entry\ E}}{\text{Total\ number\ of\ terms\ in\ entry\ E}}$. 
The semantic ($S_E$) of a syslog entry $E$ is defined as $\frac{\text{Number\ of\ terms\ with\ semantic\ in\ entry\ E}}{\text{Total\ number\ of\ terms\ in\ entry\ E}}$.
Calculating these properties for the sample syslog entry $E_1$ from Table~\ref{tab:syslog_sample}, with the information from Table~\ref{tab:sense-semantic}, results in: $N_{E{_1}}=\frac{4}{6}$
and $S_{E{_1}}=\frac{4}{6}$, respectively.
The quality of the sample syslog ($Q_{E{_1}}$) is then obtained with Equation~(\ref{quality_raw}) to be $Q_{E{_1}} = 1 * \frac{4}{6} * \frac{4}{6} * 0.75 \approx 0.33$.
The steps for performing a full anonymization with the proposed approach on the sample syslog entry $E_1$ from Table~\ref{tab:syslog_sample} are shown in Table~\ref{tab:syslog_sample_done}.

\begin{table}[h]
	\caption{Anonymization and encoding (hashing) of the sample syslog entry from Table~\ref{tab:syslog_sample}.}
	\label{tab:syslog_sample_done}
	\(\displaystyle \textbf{A)} \)
	\begin{tabular}{l|*{6}{c}}
		\textbf{Message}&Accepted&publickey&for&Siavash&from&4.3.2.1\\
		\textbf{Sensitive}&-&-&-&Y&-&Y\\
		\textbf{Semantic}&Y&Y&-&Y&-&Y\\
		\multicolumn{2}{c}{}\\
	\end{tabular}
	\(\displaystyle \mathbf{Q_{E{_1}}} = 1 * 0.67 * 0.67 * 0.75 \approx \textbf{0.33} \)\\
	\(\displaystyle \textbf{B)} \)
	\begin{tabular}{l|*{6}{c}}
		\multicolumn{2}{c}{}\\
		\hline
		\multicolumn{2}{c}{}\\
		\textbf{Anon. \#1}&Accepted&publickey&for&\#USR\#&from&4.3.2.1\\
		\textbf{Sensitive}&-&-&-&-&-&Y\\
		\textbf{Semantic}&Y&Y&-&-&-&Y\\
		\multicolumn{2}{c}{}\\
	\end{tabular}
	\(\displaystyle \mathbf{Q_{E{_1}}} = 1 * 0.83 * 0.50 * 0.75 \approx \textbf{0.31} \)\\
	\(\displaystyle \textbf{C)} \)
	\begin{tabular}{l|*{6}{c}}
		\multicolumn{2}{c}{}\\
		\hline
		\multicolumn{2}{c}{}\\
		\textbf{Anon. \#2}&Accepted&publickey&for&\#USR\#&from&\#IP4\#\\
		\textbf{Sensitive}&-&-&-&-&-&-\\
		\textbf{Semantic}&Y&Y&-&-&-&-\\
		\multicolumn{2}{c}{}\\
	\end{tabular}
	\(\displaystyle \mathbf{Q_{E{_1}}} = 1 * 1.00 * 0.33 * 0.75 \approx \textbf{0.25} \)\\
	\(\displaystyle \textbf{D)} \)
	\begin{tabular}{l|*{6}{c}}
		\multicolumn{2}{c}{}\\
		\hline
		\multicolumn{2}{c}{}\\
		\textbf{Anon. \#3}&Accepted&\#KEY\#&for&\#USR\#&from&\#IP4\#\\
		\textbf{Sensitive}&-&-&-&-&-&-\\
		\textbf{Semantic}&Y&-&-&-&-&-\\
		\multicolumn{2}{c}{}\\
	\end{tabular}
	\(\displaystyle \mathbf{Q_{E{_1}}} = 1 * 1.00 * 0.17 * 0.75 \approx \textbf{0.125} \)\\
	\(\displaystyle \textbf{E)} \)
	\begin{tabular}{l|*{6}{c}}
		\multicolumn{2}{c}{}\\
		\hline
		\multicolumn{2}{c}{}\\
		\textbf{SHAKE-128}&caa5002d&\phantom{fillup the cell}&\phantom{fillup the}& & & \\
		\textbf{Sensitive}&-&-&-&-&-&-\\
		\textbf{Semantic}&Y&-&-&-&-&-\\
		\multicolumn{2}{c}{}\\
	\end{tabular}
	\(\displaystyle \mathbf{Q_{E{_1}}} = 1 * 1.00 * 1.00 * 0.81 \approx \textbf{0.81} \)
\end{table}

The encoding algorithm used in this work is the variable length hash algorithm SHAKE-128~\cite{XOF, FIPS} with 32-bit output length adjustable based on the system requirements.
All syslog entries that follow the pattern of the sample syslog entry $E_1$: 
"\fontsize{9.3pt}{10pt}\verb|Accepted publickey for Siavash from 4.3.2.1|",
regardless of the values which they carry, after `constantification' are identical to: 
"\fontsize{9.3pt}{10pt}\verb|Accepted publickey|
\verb|for #USR# from #IP4#|".
This string is an \emph{event pattern}.
Event patterns are constant strings with a certain semantic meaning.
Replacing them with a shorter identifier does not change their meaning, as long as the identifier replacing a particular event pattern is known.
Therefore, in this work, a hashing function is used to transform event patterns from syslogs into shorter single-term identifiers.
Using a hashing function guarantees that an event pattern is always converted to an identical identifier (hash-key).
The identifier (hash-key) carries the same semantic as the event pattern, in an 8-character string.
The identifier "caa5002d" in comparison with the original string of "Accepted publickey for Siavash from 4.3.2.1" with 43 characters, represents an 81\% decrease in the string length.
Apart from shortening the syslog entries, using identifiers also reduces the number of terms in each syslog entry, which in turn, results in significant performance improvement of further processing of syslog entries.

\section{Methodology}
\label{sec:methodology}

We selected a thirteen-month collection of syslogs between February 01, 2016 and February 28, 2017, from the Taurus production HPC cluster as the source of information in the present study.
Taurus is a Linux-based parallel cluster with 2014 computing nodes. 
It employs Slurm~\cite{Yoo2003} as its batch system.
Taurus' 2014 computing nodes are divided into six islands, mainly based on their processing units type: CPUs (Intel's Haswell, Sandy Bridge, Triton, Westmere), and GPUs.

The thirteen-month collection includes syslog entries 
from all 2014 Taurus computing nodes.
The syslog daemons on the computing nodes are configured to submit syslog entries to a central node.
The central node, in turn, sends the entries to a syslog storage node.
On this storage node, daily syslog entries are accumulated according to their origin into different \emph{log files}.
Therefore, the thirteen-month collection includes 2014 system log files per day (one log file for each computing node).

\begin{figure*}[!t]
	\centering
	\includegraphics[width=\textwidth]{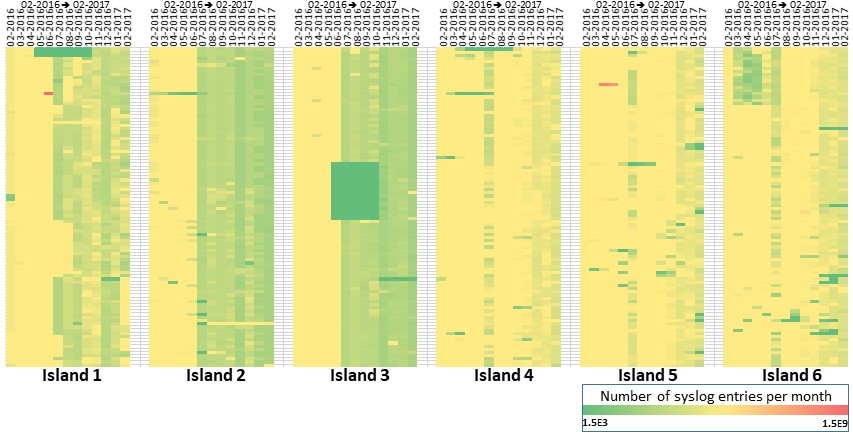}
	\caption{The number of syslog entries per node, per month. 
In each island, a column represents (from left to right) a month between February 2016 and March 2017. 
A row denotes a computing node in each island.  
The intersection between rows (nodes) and columns (months) is called a cell.
    The heat color within a cell illustrates the relative number of syslog entries for a particular node in a specific month.}
	\label{fig:syslog-entries}
\end{figure*}

The number of syslog entries generated by a computing node per day depends on various factors, including system updates and node failures.
For the thirteen-month period of this study, approximately $984.26 GiB$ of syslogs were collected, which comprise 8.6 billion syslog entries.
Fig.~\ref{fig:syslog-entries} illustrates the distribution of syslog entries among the first 100 nodes of each island.
A row in an island indicates a node and a column represents a full month period between February 2016 and March 2017.

\begin{figure}[!t]
	\centering
	\includegraphics[width=0.5\textwidth]{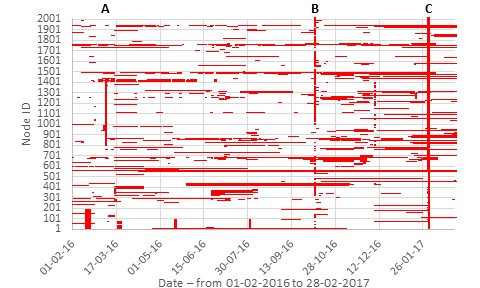}
	\caption{Illustration of the syslog entries collection gaps. 
    Approximately $3\%$ of the syslog entries collected over thirteen months were not correctly recorded. 
    The data loss occurred at three distinct points in time, identified as three vertical lines. 
    The causes of this data loss are 
    (A)~scheduled maintenance, 
    (B)~reaction of automatic overheating protection mechanism, and 
    (C)~failure of the central syslog collection node.}
	\label{fig:gaps}
\end{figure}

Various causes, such as scheduled maintenance or node failures, are responsible for a certain percentage of errors during the collection of syslog entries.
The \emph{completeness} of the syslog collection process can be measured by considering the presence of a \emph{log file} as the indicator of the gathering of syslog entries from a particular node on a given day.
Based on this definition, the syslog collection \emph{completeness} for the specific time interval in this work is $97\%$.
The red lines in Fig.~\ref{fig:gaps} indicate the $3\%$ of \emph{missed} (uncollected) syslogs.

Most of the missing $3\%$ syslog entries have been lost over the course of three days, marked at the top of Fig.~\ref{fig:gaps} with letters A, B, and C.
The reasons for their occurrence was (A)~scheduled maintenance, (B)~reaction of automatic overheating protection mechanism, and (C)~failure of the central syslog collection node.
\section{Results}
\label{sec:results}

The proposed anonymization approach has been applied to a thirteen-month collection of Taurus syslog entries.
During this process, the sensitivity and semantic of each term needed to be identified based upon the policies of Taurus HPC cluster.
According to the user privacy and data protection act of the \emph{Center for Information Services and High Performance Computing}, at the Technical University of Dresden (TUD), Germany, HPC system usage information may be anonymously collected from the users and shared with research partners.
This information includes, yet is not limited to, various metrics about processors, networks, storage systems, and power supplies~\cite{privacy}.
Other types of information are processed according to the \emph{IT}~\cite{IT} and \emph{identity management}~\cite{identity} regulations of TUD.
Based on these regulations, certain data are considered sensitive and must, therefore, remain confidential~\cite{terms}.
To the best of our knowledge, the information in Table~\ref{tab:zih-sensitive} captures the data sensitivity according to the TUD privacy regulation in force.
From this Table~\ref{tab:zih-sensitive}, one can note that certain syslog entry terms, such as node names or port numbers, may remain unchanged.
The use of the proposed approach on 8.6 billion syslog entries from Taurus of an uncompressed size of $985$ GiB, according to the TUD privacy regulations revealed seven facts.

\begin{table}[h]
	\caption{Syslog entry sensitivity according to the TUD privacy regulations}
	\label{tab:zih-sensitive}
	\begin{tabular}{l@{}|c|c}
		\textbf{Term}&\textbf{Sensitivity}&\textbf{Severity}\\
		\hline
		\fontsize{7pt}{10pt}\Verb|Surname| &Y &10 \\
		\fontsize{7pt}{10pt}\Verb|Firstname| &Y&10 \\
		\fontsize{7pt}{10pt}\Verb|Title| &Y &10 \\
		\fontsize{7pt}{10pt}\Verb|User type (employee, student, guest)| &Y &10 \\
		\fontsize{7pt}{10pt}\Verb|User name| &Y &10 \\
		\fontsize{7pt}{10pt}\Verb|Password| &Y &10 \\
		\fontsize{7pt}{10pt}\Verb|Login status (active, disabled)| &Y &10 \\
		\fontsize{7pt}{10pt}\Verb|User ID (identification of Unix users)| &Y &10 \\
		\fontsize{7pt}{10pt}\Verb|Home (Path to home directory)| &Y &10 \\
		\fontsize{7pt}{10pt}\Verb|Shell (default shell)| &Y &10 \\
		\fontsize{7pt}{10pt}\Verb|Group ID (belonging to Unix groups)| &Y &10 \\
		\fontsize{7pt}{10pt}\Verb|Mail addresses (TUD addresses)| &Y &10 \\
		\fontsize{7pt}{10pt}\Verb|IP Address| &Y &08 \\
		\fontsize{7pt}{10pt}\Verb|Port Number| &N  &00 \\
		\fontsize{7pt}{10pt}\Verb|Node Name|  &N  &00 \\
		\fontsize{7pt}{10pt}\Verb|Node ID|    &N  &00 \\
		\fontsize{7pt}{10pt}\Verb|Public Key| &Y &08 \\
		\fontsize{7pt}{10pt}\Verb|App Name|   &N  &00 \\
		\fontsize{7pt}{10pt}\Verb|Path / URL| &Y &01 \\
	\end{tabular}
\end{table}

\begin{figure}[!t]
	\centering
	\includegraphics[width=0.5\textwidth]{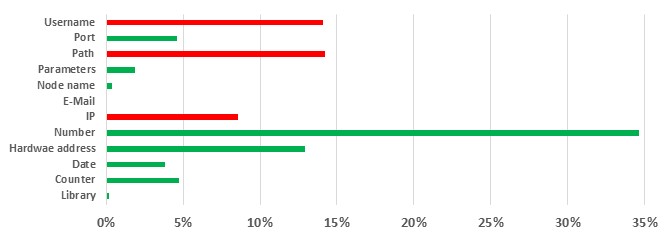}
	\caption{Percentage of sensitive and nonsensitive terms within the thirteen-month-long collection of syslogs. 
    The red bars indicate the percentage of sensitive terms while the green bars indicate the percentage of nonsensitive terms. 
    The sensitive terms sum up to approximately $35\%$ of all terms in the collection.}
	\label{fig:terms-percentage}
\end{figure}

(1)~Only approximately $35\%$ of the syslog terms are sensitive and need to be anonymized. Therefore approximately $65\%$ of terms  remained untouched (Fig.~\ref{fig:terms-percentage}).

(2)~The anonymization of sensitive terms has less than $0.5\%$ impact on syslog size reduction.

(3)~The quality of most entries degraded post-anonymization.

(4)~Approximately $2,000$ unique event patterns were discovered.

(5)~More than $90\%$ of syslog entries are based on $40$ event patterns (hereafter \emph{frequent patterns}). 
All other non-frequent event patterns together are responsible for less than $10\%$ of syslog entries.
For instance, more than $15\%$ of the syslog entries have the \fontsize{9pt}{10pt}\Verb|(#USER#) cmd (#PATH#)| pattern.

(6)~A small percentage of syslog entries (approximately $5\%$) among the non-frequent event patterns do not contain any variable terms in their original form (e.g., \fontsize{9pt}{1pt}\Verb|disabling lock debugging due to kernel taint|).

(7)~According to the current TUD privacy regulations, almost all syslog entries lose their added semantic after anonymization (e.g., "\fontsize{7.5pt}{10pt}\verb|failed password for #USER#|
\verb|from #IPv4# port 32134 ssh2|").

The only remaining useful information in these cases is the semantic of the event pattern itself.
For the above example, the useful semantic is that \fontsize{7.5pt}{10pt}\Verb|authentication via ssh failed|.
\fontsize{10pt}{10pt}
Based on the above seven observations about syslog entries on Taurus, we can state that: 
(1) The anonymized syslogs consist of approximately $90\%$ semantic-less entries (after mandatory anonymization), 
(2) Approximately $5\%$ of the entries are constant, that is they do not have any variable terms, and approximately $5\%$ are entries with semantic (and retained their useful properties even after anonymization).
Following the necessary anonymization, $(90+5)\%$ of syslog entries no longer have semantic and can be converted to hash-keys.
The $5\%$ of syslog entries which carry added semantic even after anonymization, should remain untouched.

The Table~\ref{tab:result-sample} illustrates a sample of four syslog entries in three different stages of anonymization.
The Table~\ref{tab:hash-key} is a reference to the meaning of each of the hash keys.
Together with the anonymized syslogs and according to the privacy regulations, the information in Table~\ref{tab:hash-key} may also be fully/partially published.

\begin{table}[h]
	\caption{Anonymization and hashing of four syslog entries}
	\label{tab:result-sample}
	\begin{tabular}{ll}
		 &(A) Before anonymization\\
		1&\fontsize{7.2pt}{10pt}\Verb|(siavash) cmd (/home/siavash/config.sh > output.stat)|\\
		2&\fontsize{7.2pt}{10pt}\Verb|pam_unix(sshd:session): session closed for siavash|\\
		3&\fontsize{7.2pt}{10pt}\Verb|disabling lock debugging due to kernel taint|\\
		4&\fontsize{7.2pt}{10pt}\Verb|ACPI: LAPIC (acpi_id[0x55] lapic_id[0xff] disabled)|\\
		\\
		 &(B) After anonymization, before hashing\\
		1&\fontsize{7.2pt}{10pt}\Verb|(#USER#) cmd (#PATH# > output.stat)|\\
		2&\fontsize{7.2pt}{10pt}\Verb|pam_unix(sshd:session): session closed for #USER#|\\
		3&\fontsize{7.2pt}{10pt}\Verb|disabling lock debugging due to kernel taint|\\
		4&\fontsize{7.2pt}{10pt}\Verb|ACPI: LAPIC (acpi_id[0x55] lapic_id[0xff] disabled)|\\
		\\
		 &(C) After anonymization and hashing\\
		1&\fontsize{7.2pt}{10pt}\Verb|1808e388|\\
		2&\fontsize{7.2pt}{10pt}\Verb|0964de42|\\
		3&\fontsize{7.2pt}{10pt}\Verb|59f2da35|\\
		4&\fontsize{7.2pt}{10pt}\Verb|ACPI: LAPIC (acpi_id[0x55] lapic_id[0xff] disabled)|\\
	\end{tabular}
\end{table}

\begin{table}[h]
	\caption{Hash-key reference table}
	\label{tab:hash-key}
	\begin{tabular}{l|l}
		\textbf{Hash-key}& \textbf{Meaning}\\
		\hline
		\verb|1808e388|&\fontsize{7.2pt}{10pt}\Verb|A command executed by user|\\
		\verb|0964de42|&\fontsize{7.2pt}{10pt}\Verb|A user logged out|\\
		\verb|59f2da35|&\fontsize{7.2pt}{10pt}\Verb|Disabling lock debugging due to kernel taint|\\
	\end{tabular}
\end{table}

The data in part (B) of Table~\ref{tab:result-sample} follow the main anonymization guidelines.
This fact enables their inclusion in the present work.
However, since syslog entries lengths have been reduced, the data in part (C) of Table~\ref{tab:result-sample} delivers the very same semantic as part (B), at a much smaller length.

The usefulness of the anonymized and hashed information from the thirteen-month syslog collection remains identical.
Reprocessing the results from an earlier work~\cite{SC2015}, in which the correlation of failures in Taurus was analyzed, via the new anonymization and encoding approach led to identical outcomes.
Moreover, due to single-term syslog entries, the processing time was approximately $25\%$ shorter than before.
\section{Conclusion}
\label{sec:conclusion}

System logs have widely been used in various domains, from system monitoring and performance analysis to failure prediction of different system components.
Even though system logs are mainly system dependent, having knowledge about various computing systems improves the general understanding of computing systems behavior.
However, due to the vast amount of personal data among the system log entries, users privacy concern impedes the free circulation and publication of system logs.
In this work, we examined the trade-off between sensitivity and semantic of system logs.
Since after a certain level of anonymization the semantic of system logs may be lost, keeping the semantic-less data is not the best practice.

This work introduced \emph{quality}, the system logs utility function, as a measurable parameter calculated based on nonsensitivity, semantic, reduction, and usefulness of system logs.
The goal is to maintain the quality of system log, by pushing all effective parameters to their possible limit.
This proposed approach has been applied on a thirteen-month collection of Taurus HPC cluster system logs, between from February 01, 2016 and February 28, 2017.
The proposed anonymization approach can guarantee full anonymization of syslog entries via the final encoding step.
Apart from the highest degree of anonymization, a total reduction of more than $50\%$ in system log size as well as $25\%$ performance improvement in system log analysis is achievable.

The current hashing function produces larger hash-keys than required to avoid hash-key collisions.
Fine tuning of the hashing function according to the computing system requirements, together with improving the variable term detection, are planned as future work.

\section*{Acknowledgement}
This work is in part supported by the German Research Foundation (DFG) in the Cluster of Excellence ``Center for Advancing Electronics Dresden" (cfaed).
The authors also thank Holger Mickler and the administration team of Technical University of Dresden, Germany for their support in collecting the monitoring information on the Taurus high performance computing cluster.

\section*{Disclaimer}
References to legal excerpts and regulations in this work are provided only to clarify the proposed approach and to enhance explanation.
In no event will authors of this work be liable for any incidental, indirect, consequential, or special damages of any kind, based on the information in these references. 

\bibliographystyle{IEEEtran}
\bibliography{literature}

\end{document}